\documentclass[aps,twocolumn,showpacs,superscriptaddress]{revtex4}
\usepackage{epsfig}

\def\avg#1{\langle#1\rangle}
\def\Re {\mbox{Re}}
\def\Im {\mbox{Im}}
\def\be{\begin{equation}}       \def\ee{\end{equation}}
\def\bea{\begin{eqnarray}}      \def\eea{\end{eqnarray}}
\def\PRA{Phys. Rev. A}
\def\PRB{Phys. Rev. B}
\def\PRD{Phys. Rev. D}
\def\PRL{Phys. Rev. Lett.}
\begin{document}

\title{ Exact SO(5) Symmetry in spin 3/2 fermionic system}
\author{Congjun Wu}
\affiliation{Department of Physics, McCullough Building, Stanford
University, Stanford CA 94305-4045}
\author{Jiang-ping Hu}
\affiliation{Department of Physics and Astronomy, University of
California, Los Angels, CA 90095-1547}
\author{Shou-cheng Zhang}
\affiliation{Department of Physics, McCullough Building, Stanford
University, Stanford CA 94305-4045}

\begin{abstract}
The spin 3/2 fermion models with contact interactions  have a {\it
generic} SO(5) symmetry without any fine-tuning of parameters. Its
physical consequences are discussed in both the continuum and lattice
models. A Monte-Carlo algorithm free of the sign problem at any
doping and lattice topology is designed when the singlet and
quintet interactions satisfy $U_0\le U_2\le -{3\over5} U_0~(U_0\le
0)$, thus making it possible to study different competing orders
with high numerical accuracy. This model can be accurately
realized in ultra-cold atomic systems.
\end{abstract}
\maketitle

With the rapid progress in ultra-cold atomic systems, many alkali
fermions have been cooled below Fermi temperatures \cite{demarco1999a,
truscott2001a, schreck2001a}.
All of them except $^6$Li have spins higher than 1/2 in the lowest
hyper-fine multiplets.
The spin degrees of freedom become free in the optical traps,
which has attracted interest in their effects on
Cooper pair structures and collective modes \cite{ho1999a, yip1999a}.
The proposal of the optical lattice \cite{jacksh1998a} has led to  a
tremendous progress in studying the strongly correlated bosonic lattice 
systems \cite{anderson1998a, cataliotti2001a, morsch2001a, greiner2002a,
greiner2002b}.
Recently, fermionic lattice systems are also exciting.
For example, the degenerate $^{40}$K gas
has  been prepared in a one-dimensional optical lattice
\cite{modugno2003a}.

Comprehensive analysis of symmetries is helpful in understanding the
physics in strongly correlated systems. For example, the SO(5) theory
\cite{zhang1997a} of high T$_c$ cuprates unifies the d-wave superconductivity
(SC) and antiferromagnetism (AF) orders, leading to many experimental
consequences.
The sharp neutron scattering mode can be interpreted as
the pseudo-Goldstone mode\cite{zhang1997a}. The prediction of the
antiferromagnetic vortex core \cite{arovas1997a} has also been
verified in recent experiments \cite{levi2002a}.

In this article, we focus on the symmetry properties and
corresponding consequences in the spin 3/2 system with contact interactions,
including both the continuum model with $s$-wave scattering
and the generalized lattice Hubbard model with on-site interactions.
For neutral atoms, these interactions are generally described by
two parameters in the
total spin $S_T=0,2$ channels as $g_{0,2}=4\pi\hbar^2 a_{0,2}/M$
in the continuum model with $a_{0,2}$ the corresponding $s$-wave
scattering lengths and $M$ the atom mass;  or $U_{0,2}$ in the
lattice model. Interactions in the odd total spin ($S_T=1,3$)
channels are forbidden by Pauli's exclusion principle.
Remarkably, in addition to the explicit spin SU(2) symmetry, an
enlarged SO(5) symmetry is present {\it without any fine tuning of parameters}.
In the continuum model, this symmetry has direct consequences on
the collective modes and pairing structures. In the lattice model,
exact phase boundaries of various competing phases can be
determined directly from symmetries. Because of the time-reversal
symmetry of the Kramers doublets, a Monte-Carlo algorithm free of
the notorious sign problem is designed when $ U_0\le U_2 \le -3/5
~U_0~ (U_0\le 0)$ at any filling level and lattice topology.

We start with the standard form of the spin $3/2$ Hamiltonian
of the continuum model \cite{ho1999a,yip1999a}
\bea
H&=& \int d^d{\bf r} ~ \Big\{ \sum_{\alpha=\pm 3/2, \pm 1/2}
 \psi^\dagger_\alpha({\bf r})
\big ({-\hbar^2\over 2m}\nabla^2-\mu\big) \psi_\alpha({\bf r}) \nonumber \\
&+& g_0 P_{0,0}^\dagger({\bf r}) P_{0,0}({\bf r}) +
g_2 \sum_{m=\pm2,\pm1,0} P_{2,m}^\dagger({\bf r})
P_{2,m}({\bf r})\Big \},\hspace{1cm}
\label{Ham}
\eea
with $d$ the space dimension,  $\mu$ the chemical potential and
$P^\dagger_{0,0}, P^\dagger_{2,m}$ the singlet ($S_T=0$) and quintet
($S_T=2$) pairing operators  defined through the Clebsh-Gordan
coefficient for two indistinguishable particles as $
P_{F,m}^\dagger(\bf{r})=\sum_{\alpha\beta} \avg{\frac{3}{2}
\frac{3}{2};F,m|\frac{3}{2}\frac{3}{2}\alpha\beta}
\psi^\dagger_\alpha(\bf{r}) \psi^\dagger_\beta(\bf{r}), $ where
$F=0,2$ and $m=-F,-F+1, ...,  F$.

We first construct the SO(5) algebra by introducing the five
Dirac $\Gamma^a$ $(1\le a\le 5)$ matrices
\bea
\Gamma^1=\left (
\begin{array} {cc}
0 & i I\\
-i I& 0
\end{array} \right) ,
\Gamma^{2,3,4}=\left ( \begin{array}{cc}
{\vec \sigma}& 0\\
0& {-\vec \sigma} \end{array}\right),
\Gamma^5=\left( \begin{array} {cc}
0& -I \\
-I & 0 \end{array} \right )\nonumber. \eea
Then the ten SO(5) generators are defined  as  $\Gamma^{ab}=
-{i\over 2} [ \Gamma^a, \Gamma^b] (1\le a,b\le5)$, where $I$ and
$\vec{\sigma}$ are the 2$\times$ 2 unit and  Pauli matrices.
The four-component spinor can be defined by $\psi({\bf
r})=\big ( \psi_{3\over2}({\bf r}),\psi_{1\over2}({\bf r}),
\psi_{-{1\over2}}({\bf r}),\psi_{-{3\over2}}({\bf r}) \big )^T$.
Furthermore, the bilinear operators can be classified according to
their properties under the SO(5) transformations.
The 16 bilinear operators in the particle-hole (p-h) channel can be
classified as SO(5)'s scalar, vector, and anti-symmetric tensors
(generators) as \bea &&n({\bf r})= \psi^\dagger_\alpha({\bf r})
\psi_\alpha({\bf r}), ~~ n_a({\bf r} )= \frac{1}{2}
\psi^\dagger_\alpha({\bf r}) \Gamma^a_{\alpha\beta}
\psi_\beta({\bf r}), \nonumber \\
&&L_{ab}({\bf r})=  -\frac{1}{2}\psi^\dagger_\alpha({\bf r})
\Gamma^{ab}_{\alpha\beta} \psi_\beta({\bf r}).
\eea
$L_{ab}$ and $n_a$ together form the SU(4), or isomorphically,
the SO(6) generators.
The spin SU(2) generators $J_{x,y,z}$ are expressed as
$
J_+= J_x+i~ J_y=\sqrt 3 (-L_{34}+i L_{24})+ (L_{12}+ i L_{25})
-i(L_{13}+i L_{35})$,
$J_-=J_+^\dagger$, and $J_z=-L_{23} +2~ L_{15}$.
$n$ and $n_a$ have
spin 0 and 2, and $L_{ab}$ contains both the spin 1 and 3 parts.
Pairing operators can also be organized as
SO(5) scalar and vectors through the matrix $R=\Gamma_1\Gamma_3$
\bea
\eta^\dagger({\bf r})&=&\Re \eta+ i~\Im \eta=
\frac{1}{2} \psi^\dagger_\alpha({\bf r})  R_{\alpha\beta}
\psi^\dagger_\beta({\bf r}),\nonumber\\
\chi^\dagger_a({\bf r})&=& \Re \chi_a + i~\Im \chi_a=
 -\frac{i}{2}
\psi^\dagger_\alpha({\bf r}) (\Gamma^a R)_{\alpha\beta}
\psi^\dagger_\beta ({\bf r}) \eea where $P^\dagger_{0,0}=  -
\eta^\dagger/{\sqrt 2}$, and $P^\dagger_{2,\pm2}=
(\mp\chi^\dagger_1+ i\chi^\dagger_5)/2$, $P^\dagger_{2,\pm1}=
(-\chi^\dagger_3\pm i\chi^\dagger_2)/2$, $P^\dagger_{2,0}= -i
\chi^\dagger_4/{\sqrt 2}$.
That is, $\chi^\dagger_a$ are polar
combinations of $J_z$'s eigenoperators $P^\dagger_{2,m}$.
The existence of the $R$ matrix is related to the pseudoreality 
of SO(5)'s spinor representation.
It satisfies $R^2=-1,~ R^\dagger=R^{-1}=~^t R=-R$ and $R \Gamma^a
R=-^t\Gamma^a, ~R \Gamma^{ab} R=~ ^t\Gamma^{ab}$
\cite{scalapino1998a}.
The anti-unitary time-reversal
transformation can be expressed as $T=R~ C$, where $C$ denotes
complex conjugation and $T^2=-1$. $N$, $n_a$, and $L_{ab}$
transform differently under the $T$ transformation \bea T n
T^{-1} =n,~~  T n_a T^{-1} =n_a,~~ T L_{ab} T^{-1}=-L_{ab}. \eea
 
With the above preparation, the hidden SO(5) symmetry becomes manifest.
The kinetic energy part has an explicit SU(4) symmetry which is the
unitary transformation among four spin components.
The singlet and quintet interactions are proportional to
$\eta^\dagger({\bf r}) \eta({\bf r})$ and $\chi_a^\dagger({\bf r})
\chi_a({\bf r})$ respectively, thus reducing the symmetry group from
SU(4) to SO(5). When $g_0=g_2$, the SU(4)
symmetry is restored because $\chi^\dagger_a,\eta^\dagger$
together form its 6 dimensional antisymmetrical tensor
representation. In the continuum model, interactions in other even
partial wave channels also keep the SO(5) symmetry. The odd
partial wave scattering include spin 1 and 3 channel interactions
$g_1$ and $g_3$, which together could form the 10-d adjoint representation.
of SO(5), if and only $g_1 = g_3$. However, to the leading order,
$p$-wave scattering is much weaker than the $s$-wave one for neutral
atoms,  and can thus be safely neglected.

The SO(5) symmetry implies more degeneracies in the collective excitations
in the spin 3/2 Landau fermi liquid theory,
which generally requires four fermi liquid functions
in total spin $S_T=0,1,2,3$ channels.
The SO(5) symmetry of the microscopic Hamiltonian reduces these
to three independent sets, classified according to the SO(5) scalar,
vector, and tensor channels as
\bea
f_{\alpha\beta,\gamma\delta}(p,p^\prime)&=& f_s(p,p^\prime)+
f_v(p,p^\prime) (\Gamma^a/2)_{\alpha\beta}
(\Gamma^a/2)_{\gamma\delta} \nonumber \\
&+&f_t(p,p^\prime) (\Gamma^{ab}/2)_{\alpha\beta}
(\Gamma^{ab}/2)_{\gamma\delta}. \eea In other words, the effective
interaction functions in the $S_T=1,3$ channels are exactly
identical in all orders in perturbation theory. Furthermore,
within the $s$-wave scattering approximation, the interaction
functions become constants, and are given as $f_s= (g_0+5 g_2)/16$,
$f_v= (g_0-3 g_2)/4$, $f_t= -(g_0+ g_2)/4$.
Experiments in the fermi liquid
regime can determine the four fermi liquid constants in the
$S_T=0,1,2,3$ channels separately and verify the degeneracy
between spin 1 and 3 channels.
This degeneracy appears to be accidental in Ref. \cite{yip1999a}.
It is in fact exact and protected by the generic SO(5) symmetry.

The SO(5) symmetry also enriches the Cooper pair structures.
Ref. \cite{ho1999a} showed that, in addition to the singlet pairing 
when $g_0<0$,  the spin 3/2 system energetically favors the polar 
pairing state in the quintet channel when $g_2<0$ with the order parameter
$\Delta^\dagger_{real}= \xi_1 P^\dagger_{2,0}+ \xi_2 (
P^\dagger_{2,2}+P^\dagger_{2,-2})\propto
\xi_1 \chi^\dagger_4 +\xi_2 \chi^\dagger_5$
($\xi_{1,2}$ are real).
We understand that this is only a special case of the general pairing
structures spanned by all the $\chi^\dagger_{1\sim 5}$.
The polar pairing states break the SO(5)$\otimes$ U(1) (charge) 
symmetry to $SO(4)\otimes Z_2$, and thus the Goldstone (GS)
 manifold is the quotient space $[SO(5)\otimes
U(1)]/[SO(4)\otimes Z_2]=[S^4\otimes U(1)]/Z_2$. 
Its dimension, 5, is the number of GS
modes. When both $g_{0,2}$ are positive, s-wave pairing is not
favorable. However, similarly to the spin fluctuation exchange
mediated p-wave pairing in $^3$He
\cite{anderson1973a,leggett1975a}, the spin fluctuations in the
tensor channel can induce effective attractions between two atoms
with total spin 1 and 3. This may
lead to p-wave pairing where the spin part forms the 10-d
adjoint representation of SO(5).

Now we consider the more interesting case of spin 3/2 fermions in
the optical lattice. The periodic potential is
$ V(x,y,z)=V_0 (\sin^2(k x)+\sin^2(ky)+\sin^2(kz))$
with $V_0$ the potential depth , $k=\pi/l_0$ the wavevector,
and $l_0$ the lattice constant. The hopping integral $t$  between
neighboring sites  decreases exponentially with increasing $V_0$.
Within the harmonic approximation, the parameter
$U/\Delta E\approx (\pi^2 / 2) (a_s/l_0)
(V_0/E_r)^{1/4}$,
with $U$  the repulsion of two fermions on one
site, $\Delta E$ the gap between the lowest and first excited
single particle state in one site, $a_s$ the $s$-wave scattering
length in the corresponding channel, and $E_r= \hbar^2 k^2 /2M$  the
recoil  energy. With the
typical estimate of $a_s\sim 100 a_B$ ( $a_B$ the Bohr radius),
$l_0\sim 5000$A, and $(V_0/E_r)^{1/4} \approx 1\sim 2$, we arrive
at $U/\Delta E< 0.1$. Thus this system can be approximated by
the one-band Hubbard model
\bea
\label{hmlattice1} &&H=-t\sum_{\langle ij\rangle ,\sigma} \big \{
c^\dagger_{i\sigma}
 c_{j\sigma} +h.c.\big \}-\mu \sum_{i\sigma} c^\dagger_{i\sigma} c_{i\sigma}
 \nonumber \\
&&+U_0 \sum_i P_{0,0}^\dagger(i) P_{0,0}(i) +U_2 \sum_{i,m=\pm2,\pm1,0}
P_{2,m}^\dagger(i) P_{2,m}(i). ~~~ \eea
for particle density $n\leq 4$. 
At half-filling on a bipartite lattice, $\mu$ is given by
$\mu_0=(U_0+5U_2)/4$ to ensure the particle-hole (p-h) symmetry
under the transformation $c_{i,\sigma}\rightarrow (-)^i
c^\dagger_{i,\sigma}$. The lattice fermion operators and their
continuum counterparts are related by $\psi_\alpha ({\bf r})=
c_\alpha (i)/(l_0)^{d/2}$. We use the same symbols for bilinear
fermion operators as in the continuum model.

The proof of SO(5) invariance in the continuum model applies equally well
in the lattice model at any lattice topology and at any filling level.
Eq. \ref{hmlattice1} can be conveniently rewritten in  another
manifestly SO(5) invariant form as
\bea \label{hmlattice2}
H_0&=& -t \sum_{\langle i,j \rangle}\Big \{ \psi^\dagger(i) \psi(j)
+ h.c.\Big\}  \\
H_I&=& \sum_{i, 1\le a\le 5}\Big \{
{3 U_0+5 U_2 \over 16} (n(i)-2)^2-{U_2-U_0\over 4} n_a^2(i) \Big \}
\nonumber\\
&-& (\mu -\mu_0) \sum_i n(i), \eea where the SU(4) symmetry
appears at $U_0=U_2$ as before.

\begin{table}[h]
\begin{center}
\begin{tabular}{|c|l|c|c|c|}   \hline
Phase &  order parameters  &  GS manifold & GS modes \\ \hline
A& $(-)^i L_{ab}(i)$&${SO(5) \over SO(3) \otimes SO(2)}$&6 \\ \hline
B& $ (-)^i n_a(i)$  & $SO(5)/ SO(4)\equiv S^4$ & 4 \\  \hline
C& $ \eta(i)$       & $U(1)$&1  \\ \hline
D& CDW             & /  &/  \\ \hline\hline
E& $ (-)^i n_a(i),  (-)^i L_{ab}(i)$&
   ${U(4)\over U(2)\otimes U(2)}$&  8 \\ \hline
F& $ (-)^i n_a(i), \eta(i)$& $SO(7)/SO(6)\equiv S^6$ & 6 \\ \hline
G& CDW, $\eta(i)$ &$SO(3)/SO(2)\equiv S^2$ &2 \\ \hline
H& CDW, $\chi_a(i), (-)^i L_{ab}(i)$&
${SO(7)\over SO(5)\times SO(2)}$& 10  \\ \hline
\end{tabular}
\caption{
Order parameters, the corresponding Goldstone manifolds and the number
of Goldstone modes in each phase on the bipartite lattice at half-filling.
}
\end{center}
\end{table}

The lattice Hamiltonian, Eq. \ref{hmlattice1}, contains even higher symmetries
under certain conditions.
One can construct the largest SO(8)\cite{lin1998a} algebra using
all the independent fermionic bilinear operators.
Its generators $M_{ab}~(0\le a<b\le 7)$ including $L_{ab} (1\le a<b\le 5)$
as its SO(5) sub-algebra, are  denoted as
\bea\label{so8algebra}
\hspace{-5mm}
M_{ab}=
\left( \begin{array}{cccc}
0&  \Re \chi_1~ \sim~ \Re \chi_5 & N  &   \Re \eta  \\
 &                                  &\Im \chi_1 &n_1  \\
 &        L_{ab}                    & \sim      &\sim \\
 &                                  &\Im\chi_5  &n_5 \\
 &                                  &   0       &-\Im\eta \\
 &                                  &           &0 \\
\end{array} \right), \nonumber
\eea
with $N =(n-2)/2$.
Its Casimir is a constant $C_{so8}=\sum_{0\le a<b\le 7} M^2_{ab}(i)=7$.
The global SO(8) generators are defined to be uniform in the p-h
channel as $M_{ab}=\sum_i M_{ab}(i)$ and staggered in the p-p
channel as $M_{ab}=\sum_i (-)^i M_{ab}(i)$ on the bipartite
lattice. These global generators commute with the hopping term
$[M_{ab}, H_0]=0$. On the other hand, order parameters transformed
under the SO(8) group should be staggered (uniform) in the p-h
(p-p) channel respectively. The SO(8) symmetry is always broken
by the interaction, but its subgroup symmetry, the
SO(5)$\otimes$ SU(2) and SO(7) symmetries, appear under special 
conditions as shown below.

\begin{figure}\label{phasediagram}
\vspace{-10mm}
\centering\epsfig{file=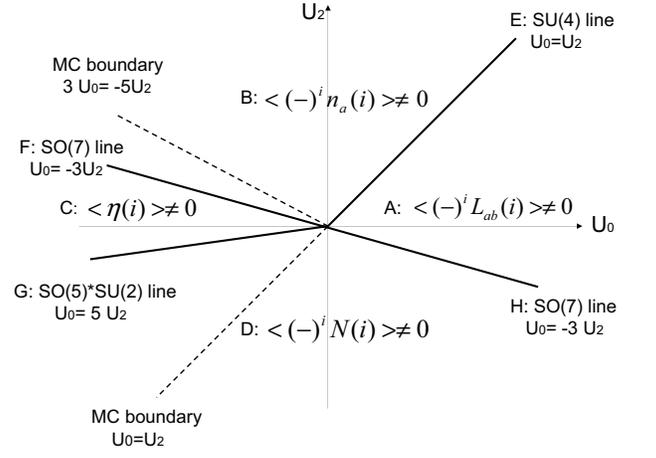,clip=1,width=75mm,angle=90}
\caption{The mean-field phase diagram at half-filling
on a bipartite lattice.
A) and B): staggered phases of the SO(5) adjoint and vector Reps;
C): the singlet superconductivity; D): CDW;
E), F), G) and H): exact phase boundaries with higher symmetries.
Between the dashed lines ($U_0\le U_2\le -3/5~ U_2$),
a Monte-Carlo algorithm free of the minus sign problem
at any filling level and lattice topology is possible.
}\label{level}
\end{figure}

At  $U_0=5~U_2$, $H_I$ can be rewritten as $ H_I=\sum_{i, 1\le a,b \le 5}
\{-U_2~L_{ab}^2(i)-(\mu-\mu_0) n(i)\}$,
using the Fierz identity $\sum_{1\le a  \le 5} L^2_{ab}(i)
+\sum_{1\le a \le 5} n_a^2(i)+5 N^2 (i)=5$.
As a generalization of the pseudospin algebra in the usual Hubbard model
\cite{yang1989a}, we construct them as $\eta^\dagger,\eta, N$.
The symmetry at half-filling is SO(5) $\otimes$ SU(2), which unifies the
charge density wave (CDW) and the singlet pairing (SP) order parameters.
Away from half-filling, this symmetry is broken
but $ \eta,\eta^\dagger$ are still eigen-operators since
$ [H, \eta^\dagger] = -(\mu-\mu_{0}) \eta^\dagger$, and $[H, \eta] =
(\mu-\mu_{0}) \eta$.

The p-h channel SO(5) $\otimes$ U(1) symmetry can also be
extended to SO(7) at $U_0=-3 U_2$ where $H_I$ can be rewritten as
$H_I= \sum_{i,0\le a <b \le 6} \Big \{ \frac {2}{3} U_2 ~
M_{ab}(i)^2 -(\mu-\mu_0) n(i)\Big \}$. The SO(7) symmetry is exact
at half-filling. Its 7-d vector Rep unifies the staggered 5-vector
and  SP order parameters. Its 21-d adjoint Rep unifies the
staggered SO(5) adjoint Rep order parameters, CDW, and quintet
pairing (QP) order parameters. Away from half-filling, QP
operators are spin 2 quasi-Goldstone operators
$[H,\chi^\dagger_{a}]=-(\mu-\mu_{0}) \chi^\dagger_{a}$ and
$[H,\chi_{a}]=(\mu-\mu_{0}) \chi_{a}$. These $\chi$ modes are just
the analogs of the $\pi$ modes in the high T$_c$
context\cite{zhang1997a}.

The above symmetry structures guide the mean field (MF) analysis of the
phase diagram.
In the weak coupling limit, the complete MF decoupling is performed in
the direct, exchange, and pairing channels.
We take the MF ansatz on the 2d square lattice
\bea
&&\avg {n_a(i)}=(-)^i \overline{n}_a ,~~~~
\avg{N(i)}=(-)^i \overline{N}, \nonumber\\
&&\avg {L_{ab}(i)} =(-)^i \overline {L}_{ab}  ~~ \avg
{\eta(i)}=\overline \eta,~~ \avg {\chi_{a}(i)}=\overline \chi_a,~
\eea
and solve it self-consistently at  half-filling to obtain
the phase diagram shown in Fig. 1. Higher symmetry lines E, F, G,
H separate phases A, B, C, D as first order phase transition
boundaries where order parameters {\it smoothly} rotate from 
one phase to another.
Symmetries on lines E, F, G, H and the order parameters are
SU(4)(adjoint Rep), SO(7)(vector Rep), SO(5)$\otimes$ SU(2)
(scalar $\otimes$ vector Rep), SO(7) (adjoint Rep) as discussed
before. Phases A and B spontaneously break the SO(5) symmetry in
the adjoint and vector Rep channels respectively. Phases C and D
have singlet pairing SC and CDW as order parameters,
respectively.
Order parameters in each phase and
corresponding GS modes are summarized in Table 1. The
effective theory is generally given by a quantum non-linear
$\sigma$ model defined on the GS manifold.

One major difficulty of Monte-Carlo simulations in fermionic systems,
the sign problem\cite{blankenbecler1981a}, is absent in the spin 3/2
model  when $U_0\le U_2\le-3/5~U_0$.
By the Hubbard-Stratonovich (HS) transformation, the partition
function can be written as below when $V=-(3U_0+5U_2)/8>0$ and
$W=(U_2-U_0)/2>0$, or equivalently $U_0\le U_2\le -3/5 ~ U_0$,
\bea Z&=& \int D n \int D n^a
\exp\Big\{ -\frac{V}{2} \int_0^\beta d\tau \sum_i n(i,\tau)^2 \nonumber \\
&-&
\frac{W}{2} \int_0^\beta d\tau \sum_{i,a} n_a^2(i,\tau) \Big \}
~~\det \Big \{ I+ B\Big \}, \nonumber
\eea
where $B= {\cal T} e^{-\int_0^\beta d\tau~ H_0+H_I(\tau)}$
and ${\cal T}$ is the time order operator.
Its discrete version is
\bea
&&B= e^{\Delta \tau H_0} e^{\Delta \tau H_I(\tau_L)} \cdot\cdot\cdot
e^{\Delta \tau H_0} e^{\Delta \tau H_I(\tau_{2})}
e^{\Delta \tau H_0} e^{\Delta \tau H_I(\tau_1)},\nonumber\\
&&H_I(\tau)= -\sum_i \psi^\dagger_\alpha(i) \psi_\alpha(i)~
\Big \{ V~ (n(i,\tau)-2) +(\mu-\mu_0) \Big \}
\nonumber \\
&&\hspace{13mm}-W\sum_{i,a} \psi^\dagger_\alpha(i)
\Gamma^a_{\alpha\beta} \psi_\beta(i)~~ n^a(i,\tau), \eea where
$\Delta \tau= \beta/L$. $I+B$ is invariant under the time-reversal
transformation: $ T(I+B) T^{-1}= I+B$. If $\lambda$ is an
eigenvalue of $I+B$ with the eigenvector $|\phi\rangle$, then
$\lambda^*$ is also an eigenvalue with the eigenvector
$T~|\phi\rangle$. From $T^2=-1$, it follows that $\avg{\phi|
T\phi}=\avg {T^2\phi|T\phi}= 0$, i.e. $|\phi\rangle$ and $T
|\phi\rangle$ are orthogonal. Thus although $I+B$ may not be
Hermitian because of the ${\cal T}$ operator, its determinant, a
product of $\lambda^*\lambda$, is always positive semi-definite.
Our proof is equally valid in the practical sampling with the discrete HS
transformation as in Ref. \cite{motome1997a}, and has been confirmed
numerically\cite{preparation}. We emphasize that this proof is valid for any
filling and lattice topology. A similar model has recently
been introduced in Ref.~\cite{assaad2002a}, where the sign problem
is also absent. However, their model keeps only the diagonal
$n_4^2$ interaction and is not spin rotationally invariant.
The valid region for the above algorithm (see
Fig. 1) includes the 5-vector phases B, SP phase C and their SO(7)
boundary, which  are analogs of the competitions between
antiferromagnetism and superconductivity in the high T$_c$
context. It would be interesting to study the doping effect, the
frustration on the triangular lattice, {\it etc}, which are
difficult at low temperatures for previous Monte-Carlo works.
Extensive numerical simulations are currently being
carried out\cite{preparation}.

Besides the alkali atoms, the trapping and cooling of the alkaline-earth
atoms are also exciting recently\cite{machholm2001,maruyama2003}.
Among these two families, $^{132}$Cs, $^9$Be, $^{135}$Ba and $^{137}$Ba are
spin 3/2 atoms. $^{132}$Cs is unstable and the $2s^2\rightarrow 2s^12p^1$
resonance of Be lies in the ultraviolet region, making them difficult
for experimental use. The resonances of the last two Ba atoms
are $6s^2\rightarrow 6s^16p^1$ at 553.7 nm \cite{he1991a},
thus they are possible candidates.
Their scattering lengths are not available now, but that of $^{138}$Ba (spin 0)
was estimated as $-41 a_B$ \cite{machholm2001},
Because the $6s$ shell of Ba is full-filled,
both the $a_0$, $a_2$ of $^{135}$Ba and $^{137}$Ba
 should have the similar value.
Considering the rapid development in this field,
we expect more and more spin 3/2 systems will be realized
experimentally, allowing us to explore the full phase diagram.

In summary, we found an exact and generic SO(5) symmetry in spin
3/2 models with local interactions. This model can be accurately
realized in cold atom systems and the theoretical predictions can
be tested experimentally by the exact relationship among the
Landau fermi liquid parameters, spin wave functions of the Cooper
pairs, exact boundaries of quantum phase transitions among various
competing states and the number of the collective modes. In the
regime where accurate Monte Carlo simulations can be carried out
without the sign problem, detailed quantitative comparisons with
experiments are possible, including the quantum phase transition
from the AF to the SC phases as a function of doping.

We thank D. P.  Arovas, A. Auerbach, B. A. Bernevig, S. Capponi, H. D. Chen, 
C. Chin, A. L. Fetter, F. Kasevich, T. K. Ng  for helpful discussions.
We especially thank E. Mukamel for carefully reading the manuscript.
This work is supported by the NSF under grant numbers DMR-9814289, and
the US Department of Energy, Office of Basic Energy Sciences under
contract DE-AC03-76SF00515. CW is also supported by the
Stanford Graduate Fellowship program.

\end{document}